\documentstyle[11pt]{article}
\def\a{\alpha}

\def\d{\delta}
\def\D{\Delta}
\def\e{\epsilon}

\def\F{\Phi}

\def\g{\gamma}
\def\G{\Gamma}

\def\k{\kappa}

\def\L{\Lambda}
\def\m{\mu}
\def\n{\nu}

\def\p{\pi}

\def\q{\theta}

\def\t{\tau}

\def\fr{\frac}
\def\ba{\begin{array}}
\def\ea{\end{array}}
\def\bz{\begin{equation}}
\def\ez{\end{equation}}
\def\by{\begin{eqnarray}}
\def\ey{\end{eqnarray}}

\def\nn{\nonumber}
\newcommand{\ls}[1]
   {\dimen0=\fontdimen6\the\font 
    \lineskip=#1\dimen0
    \advance\lineskip.5\fontdimen5\the\font
    \advance\lineskip-\dimen0
    \lineskiplimit=.9\lineskip
    \baselineskip=\lineskip
    \advance\baselineskip\dimen0
    \normallineskip\lineskip
    \normallineskiplimit\lineskiplimit
    \normalbaselineskip\baselineskip
    \ignorespaces}

\def\oldreffmt#1{\rlap{[#1]} \hbox to 2\parindent{}}

\def\figfmt#1{\rlap{Figure {#1}} \hbox to 1in{}}

\def\VEV#1{\left\langle #1\right\rangle}

\def\slash#1{#1\!\!\!/\!\,\,}
\def\beq{\begin{equation}}
\def\eeq{\end{equation}}
\def\bea{\begin{eqnarray}}
\def\eea{\end{eqnarray}}
\def\half{\frac{1}{2}}

\def\bq{\begin{quote}}
\def\eq{\end{quote}}

\def\half{\frac{1}{2}}

\def \gta {\mathrel{\vcenter
     {\hbox{$>$}\nointerlineskip\hbox{$\sim$}}}}

\newcommand{\be}{\begin{equation}}
\newcommand{\ee}{\end{equation}}
\newcommand{\bdm}{\begin{displaymath}}
\newcommand{\edm}{\end{displaymath}}
\def\simlt{\mathrel{\lower2.5pt\vbox{\lineskip=0pt\baselineskip=0pt
           \hbox{$<$}\hbox{$\sim$}}}}
\def\simgt{\mathrel{\lower2.5pt\vbox{\lineskip=0pt\baselineskip=0pt
           \hbox{$>$}\hbox{$\sim$}}}}
\catcode`@=11
\newcount\@tempcntc
\def\@citex[#1]#2{\if@filesw\immediate\write\@auxout{\string\citation{#2}}\fi
  \@tempcnta\z@\@tempcntb\m@ne\def\@citea{}\@cite{\@for\@citeb:=#2\do
    {\@ifundefined
       {b@\@citeb}{\@citeo\@tempcntb\m@ne\@citea\def\@citea{,}{\bf ?}\@warning
       {Citation `\@citeb' on page \thepage \space undefined}}%
    {\setbox\z@\hbox{\global\@tempcntc0\csname b@\@citeb\endcsname\relax}%
     \ifnum\@tempcntc=\z@ \@citeo\@tempcntb\m@ne
       \@citea\def\@citea{,}\hbox{\csname b@\@citeb\endcsname}%
     \else
      \advance\@tempcntb\@ne
      \ifnum\@tempcntb=\@tempcntc
      \else\advance\@tempcntb\m@ne\@citeo
      \@tempcnta\@tempcntc\@tempcntb\@tempcntc\fi\fi}}\@citeo}{#1}}
\def\@citeo{\ifnum\@tempcnta>\@tempcntb\else\@citea\def\@citea{,}%
  \ifnum\@tempcnta=\@tempcntb\the\@tempcnta\else
   {\advance\@tempcnta\@ne\ifnum\@tempcnta=\@tempcntb \else \def\@citea{--}\fi
    \advance\@tempcnta\m@ne\the\@tempcnta\@citea\the\@tempcntb}\fi\fi}
\catcode`@=12

\def\be{\begin{equation}} 
\def\ee{\end{equation}} 
\def\bea{\begin{eqnarray}} 
\def\eea{\end{eqnarray}} 

\begin{document} 
\begin{flushright}
 FERMILAB-Conf-97/032-T 
\end{flushright}
 
\vskip 0.2in

\begin{center}
{\bf \Large TOPCOLOR}\footnote{Invited Talk presented at
Strongly Coupled Gauge Theories `96, Nagoya, Nov. 1996}\\
\vskip 0.2in 
CHRISTOPHER T. HILL \\
\vskip 0.2in
Fermi National Accelerator Laboratory \\
P.O. Box 500, Batavia, Illinois, 60510, USA\\ 
\end{center} 
\vskip 0.2in
\begin{quote}
 We review a class of 
dynamical models in which top condensation
occurs at the weak scale, 
giving rise to the large 
top quark mass
and other phenomena. This  
typically requires a
color embedding, $SU(3)_c 
\rightarrow SU(3)_1\times
SU(3)_2$, ergo ``Topcolor.''
 Topcolor suggests a novel 
route to technicolor models in which
sequential quarks condense 
under the Topcolor interaction
to break electroweak symmetries.  
\end{quote}

\section{Top Quark Condensation} 
\subsection{Preliminary} 
 
The top quark has arrived 
with a mass of $\sim 175$
GeV.
This is remarkably large in 
comparison to all other
known fundamental particles, 
and coincidently equals the natural
scale of electroweak 
interactions, 
$1/\sqrt{2\sqrt{2}G_F}$. 
The top quark mass may be 
dynamically
generated, analogous to a 
constituent quark mass
in QCD. Then the relevant 
question becomes:
``Are there new strong interactions
that are revealing themselves 
through the
large top mass?''

If the answer is no, then the
most sensible scenario for physics 
beyond the standard
model is something like the minimal 
supersymmetric
standard model (MSSM). In this case the 
top quark mass is 
given by the infra--red renormalization 
group fixed point
\cite{RGFP}, and
many of the successful predictions of
 unification
are tied to this fact \cite{Carena}. 
 By itself, 
this is remarkable, because
the infra--red renormalization 
group fixed point is really
a consequence of the approximate 
scale invariance of the 
unified theory over the range 
of scales defining a desert,
i.e., trace anomalies which 
violate scale invariance are
proportional to beta--functions, 
and the fixed point 
corresponds to an approximate 
vanishing of the top--Yukawa
beta function.  This scale 
invariance must occur if
there is a gauge hierarchy.
{\em The top quark appears to be the only 
known particle which yields a 
nontrivial vanishing
of
its Higgs--Yukawa beta--function, 
ergo a nontrivial 
realization of
the hierarchical scale invariance, 
in the context of the MSSM.}

On the other hand, if the answer is yes, 
then there are potentially new effects 
that are necessarily (1) strongly coupled
and (2) generational, which will show
up in studies of both top and bottom, and possibly
other rare processes.  Naturalness
suggests that the scale of new dynamics 
must not be too far beyond
the top quark mass.  
Indeed, the third generation is now being studied 
extensively at 
the high energies of the Fermilab Tevatron, and LEP. 
This is the first expedition
into the multi--hundred GEV scale, and the possibility 
of new
physics emerging in unexpected ways is not precluded. 
By ``new strong dynamics in the top system''
we typically mean some kind of top 
quark condensation,
i.e., Cooper pairing of $\bar{t}_L$ 
and $t_R$ to
form a vacuum condensate. The original
top quark condensation models 
tried to identify 
all of the electroweak symmetry 
breaking 
ESB with the formation of a 
dynamical top quark mass \cite{BHL}.  
Top condensation models must either 
allow the scale of new dynamics, 
$\Lambda >>m_t$ with 
drastic 
fine-tuning, in which case the top mass 
is again controlled
by the renormalization group fixed point 
\cite{RGFP},
or invoke new dynamical mechanisms to 
try to obtain
a natural scheme.  A partial 
contribution to
ESB by top condensation is therefore 
a logical
possibility. \cite{Miransky} \cite{TC2}

\subsection{Models with Strong $U(1)$ 
Tilters (Topcolor I). }

We consider the possibility that the
top quark mass is large because
it is a combination
of a {\em dynamical condensate
component}, $(1-\epsilon)m_t$,
generated by a new strong dynamics,
together with a small {\em fundamental component},
$\epsilon m_t$, i.e, $\epsilon<<1$, generated by an
extended Technicolor\cite{TC} (ETC) or Higgs sector.
The new strong dynamics is assumed to
be chiral--critically strong but
spontaneously broken, perhaps by TC itself, at the
scale $\sim 1$ TeV, and it is
coupled preferentially to the third
generation.
The new strong dynamics therefore
occurs primarily in interactions
that involve $\overline{t}t\overline{t}t$,
$\overline{t}t\overline{b}b$, and
$\overline{b}b\overline{b}b$,
while the ETC
interactions of the form $\overline{t}t\overline{Q}Q$, where
$Q$ is a Techniquark,
are relatively
feeble.

Our basic assumptions
leave little freedom of choice in
the new dynamics. We assume a new
class of Technicolor models
incorporating ``Topcolor''.
In Topcolor~I the dynamics at the $\sim 1$ TeV scale
involves the following structure
(or a generalization thereof):\cite{TC2} \cite{TC22} 
\cite{TC23}
\beq
SU(3)_1\times SU(3)_2
\times U(1)_{Y1}\times U(1)_{Y2}
\times SU(2)_L \rightarrow
SU(3)_{QCD}\times U(1)_{EM}
\eeq
where
$SU(3)_1\times U(1)_{Y1}$
($SU(3)_2\times U(1)_{Y2}$)
generally couples preferentially
to the third (first and
second) generations.  The
$U(1)_{Yi}$
are just strongly rescaled
versions of
electroweak  $U(1)_{Y}$.
We remark that employing a new $SU(2)_{L,R}$
strong interaction in the third generation is also 
thinkable, but may
be problematic due to potentially large instanton 
effects that
violate $B+L$.  We will not explore this latter 
possibility
further.

The fermions are then assigned
$(SU(3)_1, SU(3)_2, {Y_1}, {Y_2}$) quantum numbers 
in the following
way:
\bea
(t,b)_L \;\;   &\sim  & (3,1,{1}/{3},0) \qquad \qquad
(t,b)_R \sim \left(3,1,({4}/{3},-{2}/{3}),0\right) 
\\ \nonumber
(\nu_\tau,\tau)_L &\sim & (1,1,-1,0) \qquad \qquad
\tau_R \sim \left(1,1,-2,0\right) \\ \nonumber
  & & \\ \nonumber
(u,d)_L,\;\;  (c,s)_L & \sim & (1,3,0,{1}/{3}) 
\qquad \qquad
(u,d)_R, \;\; (c,s)_R \sim \left(1,3,0,({4}/{3},
-{2}/{3})\right) \\  
\nonumber
(\nu, \ell)_L\;\; \ell = e,\mu & \sim & (1,1,0,-1) 
\qquad \qquad
\ell_R \sim \left(1,1,0,-2\right)
\eea
Topcolor must be broken, which we will assume is
accomplished through an (effective) scalar field:
\beq
\Phi \sim  (3,\bar{3}, y, -y) \label{phi_q}
\eeq
When $\Phi$ develops a VEV,
it produces the  simultaneous symmetry breaking
\beq
SU(3)_1\times SU(3)_2 \rightarrow
SU(3)_{QCD}\qquad
\makebox{ and}
\qquad
U(1)_{Y1}\times U(1)_{Y2}
\rightarrow  U(1)_{Y}
\label{sym_bre}
\eeq
The choice of $y$ will be specified below.

$SU(3)_1\times U(1)_{Y1}$ is assumed to be
strong enough to form
chiral condensates which will
naturally be tilted in the top
quark direction by the $U(1)_{Y1}$ couplings.
The theory is assumed to spontaneously break down 
to ordinary QCD
$\times U(1)_{Y}$ at a scale of $\sim 1$~TeV, 
before it becomes confining.
The isospin splitting that permits the formation 
of a $\VEV{\overline{t}t}$
condensate but disallows the $\VEV{\overline{b}b}$ 
condensate is due to the
$U(1)_{Yi}$ couplings. Since they are both larger 
than the ordinary hypercharge
gauge coupling, no significant fine--tuning is 
needed 
in principle to achieve this symmetry
breaking pattern.
The $b$--quark mass in this scheme
is then an interesting issue,
arising from a combination
of ETC effects and instantons
in  $SU(3)_1$. The $\theta$--term
in $SU(3)_1$ may manifest itself as
the CP--violating phase in the CKM matrix.
Above all, the new spectroscopy
of such a system
should begin to materialize
indirectly
in the third generation
(e.g., in $Z\rightarrow \overline{b}b$)
or perhaps at the Tevatron in top
and bottom quark production.
The symmetry breaking pattern outlined above will 
generically give rise to
three (pseudo)--Nambu--Goldstone bosons 
$\tilde{\pi}^a$,
or``top-pions'', near the top mass scale. 
If the Topcolor scale is of the order
of 1~TeV, the top-pions will
have a decay constant
of $f_\pi  \approx 50$ GeV, and a
strong coupling
given by a Goldberger--Treiman
relation,
$g_{tb\pi} \approx m_t/\sqrt{2}f_\pi\approx 2.5$,
potentially
observable in
$\tilde{\pi}^+\rightarrow t + \overline{b}$
if $m_{\tilde{\pi}} > m_t + m_b$.

We assume that ESB can be primarily driven
by a Higgs sector or Technicolor, with gauge 
group $G_{TC}$.\cite{Lane}
Technicolor can also provide
condensates which generate the
breaking of Topcolor
to QCD and $U(1)_Y$, although this can also 
be done by a Higgs field.
The coupling constants (gauge
fields) of
$SU(3)_1\times SU(3)_2$  are
respectively
$h_1$ and $h_2$ ($A^A_{1\mu}$
and $A^A_{2\mu}$)
while for $U(1)_{Y1}\times U(1)_{Y2}$
they
are respectively  ${q}_1$ and $q_2$,
$(B_{1\mu}, B_{2\mu})$.
The $U(1)_{Yi}$ fermion couplings are
then $q_i\frac{Yi}{2}$, where $Y1, Y2$
are the charges of the fermions under $U(1)_{Y1}, 
U(1)_{Y2}$ respectively.
A $(3,\overline{3})\times
(y,-y)$
Techni--condensate (or Higgs field) breaks 
$SU(3)_1\times SU(3)_2
\times U(1)_{Y1}\times U(1)_{Y2}
\rightarrow SU(3)_{QCD}\times  U(1)_Y$
at a scale
$\Lambda \gta 240$ GeV, or it
fully breaks $SU(3)_1\times SU(3)_2
\times U(1)_{Y1}\times U(1)_{Y2}\times SU(2)_L
\rightarrow SU(3)_{QCD}\times  U(1)_{EM}$
at the scale $\Lambda_{TC}= 240$ GeV.
Either scenario typically
leaves a {\em residual global symmetry},
$SU(3)'\times U(1)'$,
implying a degenerate, massive
color octet of ``colorons,'' $B_\mu^A$,
and a singlet heavy
$Z'_{\mu}$.  The gluon $A_\mu^A$
and coloron $B_\mu^A$ (the SM $U(1)_Y$
field
$B_\mu$ and the $U(1)'$ field
$Z'_\mu$), are then defined by
orthogonal
rotations with mixing angle
$\theta$ ($\theta'$):
\bea
& &
h_1\sin\theta = g_3;\qquad
h_2\cos\theta = g_3;\qquad
\cot\theta = h_1/h_2;\qquad
\frac{1}{g_3^2} = \frac{1}{h_1^2} +
 \frac{1}{h_2^2} ;
\nonumber \\
& &
q_1\sin\theta' = g_1;\qquad
q_2\cos\theta' = g_1;\qquad
\cot\theta' = q_1/q_2;\qquad \nonumber \\
& &
\frac{1}{g_1^2} = \frac{1}{q_1^2} +
 \frac{1}{q_2^2} ;
\eea
and $g_3$ ($g_1$) is the QCD ($U(1)_Y$)
coupling constant at $\Lambda_{TC}$.
adjusted to unity
abelian $U(1)$ couplings is  
We ultimately demand $\cot\theta \gg 1$
and  $\cot\theta' \gg 1$
to select the top quark direction for
 condensation.
The masses of the degenerate octet of
colorons and $Z'$ are given
by $M_B\approx g_3\Lambda/\sin\theta\cos\theta$
$ M_{Z'} \approx g_1\Lambda/\sin\theta'\cos\theta'$.
The usual QCD gluonic ($U(1)_Y$ electroweak)
interactions
are obtained for any quarks that carry
either $SU(3)_1$
or $SU(3)_2$ triplet quantum numbers
(or $U(1)_{Yi}$ charges).

The coupling of the new heavy bosons $Z'$ 
and $B^A$ to fermions
is then given by
\begin{equation}\label{lzb}
{\cal L}_{Z'}=g_1\cot\theta' (Z'\cdot J_{Z'}) 
\qquad
{\cal L}_{B}=g_3\cot\theta (B^A\cdot J^A_{B})
\end{equation}
where the currents $J_{Z'}$ and $J_B$ in 
general involve all three
generations of fermions
\begin{equation}\label{jzb123}
J_{Z'}=J_{Z',1}+J_{Z',2}+J_{Z',3}\qquad
J_B=J_{B,1}+J_{B,2}+J_{B,3}
\end{equation}
For the third generation the currents 
read explicitly
(in a weak eigenbasis):
\begin{eqnarray}\label{jz3}
J^\mu_{Z',3} &=& \frac{1}{6}\bar t_L\gamma^\mu t_L+
\frac{1}{6}\bar b_L\gamma^\mu b_L+
\frac{2}{3}\bar t_R\gamma^\mu t_R
-\frac{1}{3}\bar b_R\gamma^\mu b_R \\ \nonumber
& & -\frac{1}{2}\bar\nu_{\tau L}
\gamma^\mu \nu_{\tau L}
-\frac{1}{2}\bar\tau_L\gamma^\mu\tau_L-
\bar\tau_R\gamma^\mu\tau_R
\end{eqnarray}
\begin{equation}\label{jb3}
J^{A,\mu}_{B,3}=\bar t\gamma^\mu
\frac{\lambda^A}{2}t+
\bar b\gamma^\mu\frac{\lambda^A}{2}b
\end{equation}
where $\lambda^A$ is a Gell-Mann matrix 
acting on color indices.
For the first two generations the 
expressions are similar, except for
a suppression factor of $-\tan^2\theta'$ 
($-\tan^2\theta$)
\begin{equation}\label{jz2}
J^\mu_{Z',2}=-\tan^2\theta'\left(
\frac{1}{6}\bar c_L\gamma^\mu c_L+
\frac{1}{6}\bar s_L\gamma^\mu s_L 
+\ldots \right)
\end{equation}
\begin{equation}\label{jb2}
J^\mu_{B,2}=-\tan^2\theta\left(\bar  
c\gamma^\mu\frac{\lambda^A}{2}c+
\bar s\gamma^\mu\frac{\lambda^A}{2}s 
\right)
\end{equation}
with corresponding formulae applying to 
the first generation.
Integrating out the heavy bosons $Z'$ and 
$B$, these couplings give  
rise
to effective low energy four fermion 
interactions. 
The effective Topcolor interaction
of the third generation takes the form:
\bea
{\cal{L}}'_{TopC} & =  & 
-\frac{2\pi\kappa}{M_B^2}\left(
\bar{t}\gamma_\mu \frac{\lambda^A}{2} t +
 \bar{b}\gamma_\mu \frac{\lambda^A}{2} b
\right)
\left(
\bar{t}\gamma^\mu \frac{\lambda^A}{2} t +
 \bar{b}\gamma^\mu \frac{\lambda^A}{2} b
\right) .\label{topc_in}
\eea
where
\begin{equation}\label{kk1def}
\kappa_1=\frac{g^2_1\cot^2\theta'}{4\pi} \qquad
\kappa=\frac{g^2_3\cot^2\theta}{4\pi}
\end{equation}

This interaction is attractive in the 
color-singlet $\bar{t}t$ and $\bar{b}b$
channels and
invariant under color $SU(3)$
and $SU(2)_L\times SU(2)_R
\times U(1) \times U(1)$ where $SU(2)_R$ 
is the custodial
symmetry of the electroweak interactions.

In addition to the Topcolor interaction, 
we have the $U(1)_{Y1}$
interaction (which breaks custodial $SU(2)_R$):
\bz
{\cal{L}}'_{Y1} =   -\frac{2\pi\kappa_1}{M^2_{Z'}}\left(
\frac{1}{6}\bar{\psi}_L\gamma_\mu \psi_L +
\frac{2}{3}\bar{t}_R\gamma_\mu t_R
-\frac{1}{3}\bar{b}_R\gamma_\mu  b_R -
\frac{1}{2}\bar{\ell}_L\gamma_\mu \ell_L -
\bar{\tau}_R\gamma_\mu \tau_R \right)^2
\label{u1_in}
\ez
where $\psi_L = (t,b)_L$, $\ell_L = (\nu_\tau,\tau)_L$
and $\kappa_1$ is assumed to be $O(1)$. 
(A small value for $\kappa_1$ would
signify fine-tuning and may be 
phenomenologically undesirable.)

The attractive Topcolor interaction, 
for sufficiently large
$\kappa$, can trigger the formation of a
low energy
condensate, $\VEV{\overline{t}t + 
\overline{b}b}$, which
would break $SU(2)_L\times SU(2)_R\times  U(1)_Y
\rightarrow U(1)\times SU(2)_{c}$, 
where $SU(2)_{c}$ is
a global custodial symmetry. On the
other hand,  the $U(1)_{Y1}$ force 
is attractive in the  
${\overline{t}t}$
channel and repulsive in the ${\overline{b}b}$ 
channel.  Thus, we can
have in concert critical and subcritical values 
of the combinations:
\beq
\kappa + \frac{2\,\kappa_1}{9N_c} > \kappa_{crit} ;
\qquad
\kappa_{crit} > \kappa - \frac{\kappa_1}{9N_c};
\label{crit_con}
\eeq
Here $N_c$ is the number of colors. It should 
be mentioned that this analysis
is performed in the context of a large-$N_c$ 
approximation. The leading
isospin-breaking effects are kept even though 
they are ${\cal O}(1/N_c)$. The
critical coupling, in this approximation, is 
given by $\kappa_{crit} =
2\pi/N_c$. In what follows, we will not make 
explicit the $N_c$ dependence, but
rather take $N_c=3$. 
We would expect the cut--off for integrals
in the usual Nambu--Jona-Lasinio (NJL) gap 
equation for $SU(3)_{TopC}$
($U(1)_{Y1}$) to be $\sim M_B$ ($\sim M_{Z'}$). 
Hence, these  
relations define
criticality conditions irrespective of $M_{Z'}/M_B$.
This leads to ``tilted'' gap equations in 
which the top
quark acquires a constituent mass, while the 
$b$ quark
remains massless.  Given that both $\kappa$ 
and $ \kappa_1$
are large there is no particular fine--tuning 
occuring here,
only ``rough--tuning'' of the desired tilted 
configuration.
Of course, the NJL approximation is crude, 
but as long as the  
associated
phase transitions of the real strongly
coupled theory are approximately second order, 
analogous
rough--tuning in the full theory is possible.

\vspace{1.7in}
\noindent
This Figure appears in: G. Buchalla, G. Burdman, 
C. T. Hill, D. Kominis, 
{\em Phys. Rev. } {\bf D53} 5185, (1996).
\vspace{1.7in}
\begin{center}
{\small Fig.(1) The full phase diagram of the 
Topcolor I model.
}
\end{center}

\vspace{0.2in}

The full phase diagram of the model is shown 
in Fig.~1.\cite{Burdman}
The criticality conditions (\ref{crit_con}) 
define the allowed region  
in
the $\k_1$--$\k$ plane in the form of the 
two straight solid lines  
intersecting
at $(\k_1=0,\k=\k_{crit})$. To the left of 
these lines lies the  
symmetric
phase, in between them the region where only 
a $\VEV{\bar{t}t}$  
condensate forms
and to the right of them the phase where 
both $\VEV{\bar{t}t}$ and
$\VEV{\bar{b}b}$ condensates arise. The 
horizontal line marks the  
region above
which  $\k_1$ makes the $U(1)_{Y1}$ interaction 
strong enough to  
produce
a $\VEV{\bar{\t}\t}$ condensate. (This line 
is meant 
only as an indication, as
the fermion-bubble (large-$N_c$) approximation, 
which we use, evidently fails
for leptons.)
There is an additional constraint from
the measurement of $\G(Z\to\t^+\t^-)$, 
confining the allowed region  
to the one below the solid curve. This curve 
corresponds to a $2\sigma$
discrepancy between the
Topcolor prediction and the measured value of 
this width.
In the allowed
region a top condensate alone forms. The  
constraints favor
a strong $SU(3)_{TopC}$ coupling and a 
relatively weaker $U(1)_{Y1}$
coupling.

\subsection{~Anomaly--Free Models Without Strong $U(1)$ Tilters 
(Topcolor~II).}

The strong $U(1)$ is present in the previous 
scheme to avoid
a degenerate $\VEV{\bar{t}t}$ with $\VEV{\bar{b}b}$.  
However,
we can give a model in which there is: 
(i) a Topcolor $SU(3)$
group but (ii) no strong $U(1)$ with (iii) an 
anomaly-free
representation content. In fact the original
model of \cite{TC2} was of this form, introducing a
new quark of charge $-1/3$.  Let us 
consider a
generalization of this scheme which 
consists of the gauge structure
$SU(3)_Q\times SU(3)_1\times SU(3)_2 
\times U(1)_{Y}\times SU(2)_{L}$.
We require an additional triplet of fermions
fields $(Q_R^a)$ transforming as $(3,3,1)$
and $Q_L^{\dot{a}}$ transforming as $(3,1,3)$ under
the $SU(3)_Q\times SU(3)_1\times SU(3)_2$.

The fermions are then assigned the following 
quantum numbers
in $SU(2)\times SU(3)_Q \times SU(3)_1\times SU(3)_2\times U(1)_Y $:
\bea
(t,b)_L \;\;  (c,s)_L &\sim & (2,1,3,1) 
\qquad Y=1/3 \\ \nonumber
(t)_R &\sim & (1,1,3,1) \qquad Y=4/3;\qquad \\
\nonumber
(Q)_R &\sim  & (1,3,3,1) \qquad Y=0\\ \nonumber
  & & \\ \nonumber
(u,d)_L &\sim & (2,1,1,3) \qquad Y=1/3 \\ 
\nonumber
(u,d)_R \;\; (c,s)_R &\sim & (1,1,1,3) 
\qquad Y=(4/3,-2/3) \\  
\nonumber
(\nu, \ell)_L\;\; \ell = e,\mu,\tau &\sim & 
(2,1,1,1) \qquad Y=-1;  
\qquad
\\ \nonumber (\ell)_R &\sim &  (1,1,1,1) 
\qquad Y=-2 \\ \nonumber
b_R & \sim & (1,1,1,3) \qquad Y= 2/3;   \qquad
\\ \nonumber  (Q)_L
& \sim & (1,3,1,3) \qquad Y= 0;
\eea
 Thus, the $Q$ fields are electrically
neutral.  One can verify that this 
assignment is anomaly free.

The $SU(3)_Q$
confines and
forms a $\VEV{\bar{Q}Q}$ condensate which acts
like the $\Phi$ field and breaks the Topcolor
group down to QCD dynamically.    
We assume that $Q$ is then  
decoupled from the
low energy spectrum by its large 
constituent mass.  There is only a  
lone
$U(1)$ Nambu--Goldstone boson  
$\sim {\bar{Q}\gamma^5 Q}$
which acquires a large mass by 
$SU(3)_Q$
instantons.

\subsection{~Triangular Textures }

The texture of the fermion mass 
matrices will generally
be controlled by the symmetry 
breaking pattern of a horizontal
symmetry.  In the present case 
we are specifying a residual
Topcolor symmetry, presumably  
subsequent to some
initial breaking at some 
scale $\Lambda$, large compared 
to Topcolor, e.g., the third
generation fermions in Model~I have 
different Topcolor assignments than  
do the
second and first generation fermions. 
Thus the texture will depend
in some way upon the breaking of 
Topcolor.\cite{TC22} \cite{Lane}

Let us study a fundamental Higgs boson, 
which ultimately
breaks $SU(2)_L\times U(1)_Y$,
together with an effective field $\Phi$  
breaking Topcolor
as in eq.(\ref{sym_bre}).  We must now 
specify the full Topcolor  
charges
of these fields. As an example, under
$SU(3)_1\times SU(3)_2 \times U(1)_{Y1}
\times U(1)_{Y2}\times  
SU(2)_L$
let us choose:
\beq
\Phi \sim (3,\bar{3}, \frac{1}{3}, 
-\frac{1}{3}, 0)
\qquad
H \sim (1,1,0, -1, \half)
\eeq
The effective couplings to fermions that 
generate mass terms in the
up sector are of the form
\by
{\cal L}_{{\cal M}_U}&=&m_0 \bar{t}_Lt_R 
+c_{33}\bar{T}_Lt_R  
H\fr{det\F^\dagger}{\L^3}+
c_{32}\bar{T}_L c_R H\fr{\F}{\L} 
+ c_{31}\bar{T}_L u_R  
H\fr{\F}{\L}\nn\\
& &+c_{23}\bar{C}_L t_R H \F^\dagger  
\fr{det\F^\dagger}{\L^4}+c_{22}\bar{C}_L c_R
H + c_{21}\bar{C}_L u_R H  \label{lag_mass}\\
& & + c_{13}\bar{F}_L t_R H \F^\dagger  
\fr{det\F^\dagger}{\L^4}+c_{12}\bar{F}_L c_RH
+c_{11}\bar{F}_L u_R H  + {\rm h.c.} \nn
\ey
Here $T=(t,b)$, $C=(c,s)$ and $F=(u,d)$. 
 The mass $m_0$ is
the dynamical condensate top mass.
Furthermore $det\Phi$ is defined by
\bz
det \Phi \equiv \frac{1}{6}\e_{ijk}
\e_{lmn}\Phi_{il}\Phi_{jm}\Phi_{kn}
\label{phi_det}
\ez
where in $\Phi_{rs}$ the first(second) 
index refers to $SU(3)_1$ ($SU(3)_2$).  
The matrix elements now require factors
of $\Phi$ to connect the third with the
first or second generation color indices. 
The down quark
and lepton mass matrices are generated 
by couplings analogous to
(\ref{lag_mass}).

To see what kinds of textures can 
arise naturally,
let us assume that the ratio $\Phi/\Lambda$ 
is small, O($\epsilon$).
The field $H$ acquires a VEV of $v$.
Then the resulting mass  matrix 
is approximately triangular:
\bea
&& \left( \begin{array}{ccc}
c_{11}v &  c_{12}v & \sim 0    \cr 
 c_{21}v & c_{22}v  &\sim 0    \cr
c_{31}O(\epsilon)v & c_{32}O(\epsilon)v 
& \sim m_0 + O(\epsilon^3)v    
\cr
 \end{array}\right)\label{m_trian}
\eea
where we have kept only terms of 
$\cal O (\e)$ or larger.

This is a triangular matrix 
(up to the $c_{12}$ term).  
When it is written in the form
$U_L {\cal D} U^{\dagger}_R$ with 
$U_L$ and $U_R$ unitary and ${\cal  
D}$
positive diagonal,
there automatically result 
restrictions on $U_L$ and $U_R$.  In the  
present case,
the elements $U^{3,i}_L$ and $U^{i,3}_L$ 
are vanishing for
$i\neq 3$ , while the elements of 
$U_R$ are not constrained
by triangularity.
Analogously, in the down quark
sector $D^{i,3}_L=D^{3,i}_L=0$ for  
$i\neq 3$
with $D_{R}$ unrestricted.
The situation is reversed when 
the opposite corner elements are  
small,
which can be achieved by choosing 
$H \sim (1,1, -1, 0, \half)$.

These restrictions on the quark 
rotation matrices have important
phenomenological consequences.
For instance, in the process 
$B^0\rightarrow \overline{B^0}$ there are  
potentially
large contributions from top-pion 
and coloron exchange.  However, as  
we show in
Section~IV.(B), these contributions 
are proportional to the
product $D^{3,2}_L D^{3,2}_R$.  
The same occurs in $D^0-\bar{D}^0$  
mixing, where
the effect goes as products 
involving $U_L$ and $U_R$ off-diagonal  
elements.
Therefore, triangularity can 
naturally select
these products to be small.

Selection rules will be a general 
consequence in models where the
generations have different  gauge 
quantum numbers above some scale.
The precise selection rules depend 
upon the particular symmetry  
breaking
that occurs. This example is merely 
illustrative of the
systematic effects that can occur 
in such schemes.

\subsection{Top-pions; Instantons; The b-quark mass.}
Since the 
top condensation is a spectator 
to the TC (or Higgs) driven
ESB, there must
occur a multiplet of top-pions.
A chiral Lagrangian can be written:
\beq
L = i\overline{\psi}
\slash{\partial}\psi - m_t(
\overline{\psi}_L\Sigma P
\psi_R + h.c.) -\epsilon m_t
\overline{\psi}P\psi, \qquad
P=\left(\begin{array}{cc} 1 & 0\\ 0 & 0
\end{array}\right)
\eeq
and $\psi=(t,b)$, and $\Sigma = 
\exp(i\tilde{\pi}^a\tau^a
/\sqrt{2}f_\pi)$. Eq.(7) is 
invariant under
$\psi_L\rightarrow e^{i\theta^a\tau^a/2}\psi_L$,
$\tilde{\pi}^a\rightarrow 
\tilde{\pi}^a + 
\theta^a f_\pi/\sqrt{2}$.
Hence, the relevant currents 
are left-handed,
$j_\mu^a = \psi_L\gamma_\mu
\frac{\tau^a}{2}\psi_L$,
and $<\tilde{\pi}^a|j_\mu^b|0> 
= \frac{f_\pi}{\sqrt{2}}p_\mu 
\delta^{ab}$.  The Pagels-Stokar 
relation, eq.(1),
then follows by demanding that the 
$\tilde{\pi}^a$ kinetic
term is generated by integrating 
out the fermions.  The 
top--pion decay constant estimated 
from eq.(1) using
$\Lambda = M_{B}$ and $m_t = 175$ 
GeV is $f_\pi \approx 50$ GeV.
The couplings of the top-pions 
take the form:
\beq
 \frac{m_t}{\sqrt{2}f_\pi} 
\left[ {i}\overline{t} 
\gamma^5 t \tilde{\pi}^0 
+\frac{i}{\sqrt{2} }
\overline{t} (1-\gamma^5) b \tilde{\pi}^+ 
+ \frac{i}{\sqrt{2} }
\overline{b} (1+\gamma^5)t  \tilde{\pi}^- 
\right]
\eeq
and the coupling strength is 
governed by the relation 
$g_{bt\tilde{\pi}} \approx m_t/\sqrt{2}f_\pi$.

The small ETC mass component of 
the top quark implies that the masses
of the top-pions will depend 
upon $\epsilon$ and $\Lambda$. 
Estimating the induced top-pion 
mass from the fermion loop yields:
\beq
m_{\tilde{\pi}}^2 = 
\frac{N \epsilon m_t^2 M_B^2 }{8\pi^2 f_\pi^2} 
= \frac{\epsilon M_B^2 }{\log(M_B/m_t)}
\eeq
where the Pagels-Stokar formula 
is used for $f_\pi^2$
(with $k=0$) in the
last expression. For 
$\epsilon = (0.03,\; 0.1)$, 
$M_B\approx (1.5,\; 1.0) $ TeV, 
and $m_t=180$ GeV  this 
predicts $m_{\tilde{\pi}}= (180,\; 240)$ GeV.
The bare value of $\epsilon$ 
generated at the  ETC scale 
$\Lambda_{ETC}$, however, 
is subject to very large 
radiative enhancements by 
Topcolor and $U(1)_{Y1}$ by 
factors of order 
$(\Lambda_{ETC}/M_B)^p \sim 10^1$,
where the $p\sim O(1)$. 
Thus, we expect that
even a  bare value of 
$\epsilon_0 \sim 0.005$
can produce sizeable $m_{\tilde{\pi}} > m_t$.
Note that $\tilde{\pi}$ will generally 
receive gauge contributions 
to it's mass; these are
at most electroweak in strength, 
and therefore of
order $\sim 10$ GeV.

Top-pions can be as light as 
$\sim 150$ GeV, in
which case they would emerge as a 
detectable branching
fraction of top decay \cite{BB}.  
However, there are
dangerous effects in $Z\rightarrow 
b\bar{b}$ with
low mass top pions and decay 
constamnts as small as $\sim
60 $ GeV \cite{Burdman2}.  A 
comfortable 
phenomenological range is slightly 
larger
than our estimates, $m_{\tilde{\pi}} 
\gta 300$ GeV and
$f_\pi \gta 100$ GeV.

The $b$ quark receives 
mass contributions from ETC of $O(1)$
GeV, but also an induced mass
from instantons in $SU(3)_{1}$. 
The instanton
effective Lagrangian may be 
approximated by the
`t Hooft flavor determinant (
we place the cut-off at $M_B$):
\beq
L_{eff} = \frac{k}{M_B^2} 
e^{i\theta_1} \det 
(\overline{q}_L q_R) + h.c.
= \frac{k}{M_B^2} 
e^{i\theta_1}[(\overline{b}_L b_R)
(\overline{t}_Lt_R)
- (\overline{t}_L b_R)
(\overline{b}_Lt_R)] + h.c.
\eeq
where $\theta_1$ is the 
$SU(3)_1$
strong $CP$--violation 
phase. $\theta_1$ 
cannot be eliminated 
because
of the ETC contribution 
to the $t$ and $b$ masses.  
It 
can lead to
induced scalar couplings 
of the neutral top--pion \cite{Burdman},
and
an induced CKM CP--phase, 
however, we will presently
neglect the effects of 
$\theta_1$.

We generally expect $k\sim 1$ 
to $10^{-1}$ as in QCD.
Bosonizing in fermion bubble 
approximation
$\overline{q}^i_Lt_R \sim 
\frac{N}{8\pi^2} m_t M_B^2 
\Sigma^i_1$, where 
$\Sigma^i_j = 
\exp(i\tilde{\pi}^a\tau^a/
\sqrt{2}f_\pi)^i_j$ yields:  
\beq
L_{eff} \rightarrow  
\frac{Nk m_t}{8\pi^2}  
e^{i\theta}[(\overline{b}_L b_R)
\Sigma^1_{1} +
(\overline{t}_Lb_R)
\Sigma^2_{1} + h.c.]
\eeq
This implies an instanton 
induced $b$-quark mass:
\beq
m^\star_b \approx 
\frac{3 k m_t}{8\pi^2} \sim 6.6\; k\; GeV 
\eeq
This is not an unreasonable 
estimate of the observed $b$ 
quark mass
as we might have feared it
 would be too large. 
Expanding $\Sigma^i_j$, 
there also occur induced 
top--pion couplings to $b_R$:
\beq
\frac{m^\star_b}{\sqrt{2}f_\pi} 
( i\overline{b} \gamma^5 b \tilde{\pi}^0 
+ 
\frac{i}{\sqrt{2}}\overline{t} 
(1+\gamma^5) b \tilde{\pi}^+   
+ \frac{i}{\sqrt{2}}\overline{b}
(1-\gamma^5) t \tilde{\pi}^- )
\eeq

\section{Low Energy Observables }

We summarize some of the 
consequences of Topcolor dynamics
in low energy processes. 
This is discussed in greater
detail elsewhere.\cite{Burdman}
Potentially large FCNC arise 
when the quark
fields are rotated from their 
weak eigenbasis to their mass eigenbasis. 
In the case of Topcolor I, 
the presence of a residual $U(1)_{Y}$ interacting
strongly with the third generation 
implies that the $Z'$ will also couple
to leptons in order to cancel 
anomalies, generating contributions
to semileptonic processes. 
In Topcolor~II the induced
four--fermion interactions 
remain nonleptonic, hence the semileptonic
processes are model dependent. 

For quark field rotations are 
involved we must choose 
an anzatz for  $U_{L,R}$ and $D_{L,R}$. A 
conservative choice is to take 
the squared root of the CKM 
matrix indicative of the order
of magnitude of the effects.  
However,  triangular textures 
imply the vanishing of some  
of the non--diagonal 
elements, suppressing large 
contributions 
to  mixing.  Presently we summarize 
only the largest
potential effects.\cite{Burdman}

\subsection{ ~Semileptonic Processes}

The couplings of the Topcolor I $Z'$ 
to quarks are given by
eq.(15).  In going to the mass eigenbasis 
the quark fields are rotated by the matrices
$U_L$, $U_R$ (for the up-type left and right 
handed quarks) and $D_L$, $D_R$
(for the down-type left and right 
handed quarks).
We make the replacement
\bz
b_L\to D_{L}^{bb}b_L + D_{L}^{bs}s_L 
+ D_{L}^{bd}d_L \label{bl_rot}
\ez
and analogously for $b_R$. Thus there
will be induced FCNC interactions
from eq.(15).

\vskip 0.1in

\noindent
{\em\bf 1.~ $B_s\to \ell^+\ell^-$}
\vskip 0.1in
\noindent A flavor changing  
$b-s$ coupling  is induced:
\bz
{\cal L}_{bsZ'}=-\frac{g_1}{2}\cot\q '
Z'^{\mu}\left\{\fr{1}{3}D_{L}^{bb}
D_{L}^{bs*}\bar{s}_L\g_\m b_L -
\fr{2}{3}D_{R}^{bb}D_{R}^{bs*}\bar{s}_R
\g_\m b_R  +{\rm h.c.} \right\}
\label{bsz},
\ez
where the coupling to leptons is:
\bz
{\cal L}_{\ell\ell Z'}=
-\frac{g_1}{2} X(\q ')
Z^{'\mu}\left\{(-1)\bar{\ell}_L\g_\m\ell_L
+(-2)\bar{\ell}_R\g_\m\ell_R +{\rm
h.c.} \right\} \label{llz}
\ez
where $X(\q ')=\cot\q '$ for $\ell=\tau$, 
and $X(\q ')=\tan\q '$ otherwise.

Assuming only the Topcolor contribution, 
the $B_s$ width is given by
\bz
\Gamma(B_s\to\ell^+\ell^-)=\frac{1}{4608\,\pi} 
f_{B_s}^{2}
m_{B_s}m_{\ell}^{2}\,\sqrt{1-
\frac{4m_{\ell}^{2}}{m_{B_s}^{2}}}
\,\delta_{bs}^{2}\cot^2\q ' X^2(\q ')
\left(\frac{g_1}{M_{Z'}}\right)^4
\label{gamma_ll}
\ez
where we've defined:
\bz
\delta_{bs}=D_{L}^{bb}D_{L}^{bs*}
+2D_{R}^{bb}D_{R}^{bs*} \label{delta}
\ez
and $B_s$ is the $b\bar{s}$ meson 
decay constant.

The most significant effect occurs 
in $B_s\to \tau^+\tau^-$, 
given that the
strong $U(1)_{Y1}$ couples to 
the $\t$ lepton. 
 This implies that $X(\q ')=
\cot \q '$. Using
$f_{B_s}=0.200$~GeV and 
defining:
\bz
\kappa_1=\frac{g_{1}^{2} 
\cot^2\q '}{4\pi} \label{k1_def}
\ez
one gets:
\bz
\Gamma(B_s\to \tau^+\tau^-)= 
6\times 10^{-3}
\;\delta_{bs}^{2}\;
\frac{k_{1}^{2}}{M_{Z'}^{4}} 
\;{\rm GeV}^5 \label{gamma_tt}
\ez
For the neutral mixing factors 
we make use of the CKM squared 
root ansatz:
($\sqrt{\rm CKM})$
\bz
D_{L,R}^{bs}\approx 1/2 |V_{cb}| 
\label{sqran}
\ez
which in this case is rather 
general given that (\ref{delta}) involves 
a sum of left and right 
contributions and, for instance,
it will not vanish when the 
textures are triangular as in (\ref{m_trian}). 
We still have the freedom of 
the relative sign between the elements of  $D_{L}$
and $D_{R}$ in (\ref{delta}). 
This introduces an uncertainty of a factor of $3$
in the amplitude.
Taking  $k_1\simeq {\cal O}(1)$ we get
\bz
BR(B_s\to \tau^+\tau^-)\approx
\left\{
\begin{array}{c}
 1\;(0.1)\times 10^{-3}\;\;\; 
{\rm for }\;\; m_{Z'}=\; 500 {\rm GeV} \\
\; 6\;(0.7)\times 10^{-5}  \;\;\; 
{\rm for }\;\; m_{Z'}=1000 {\rm GeV}
\end{array}
\right.
\ez
where we have used the positive 
(negative) relative sign in (\ref{delta}).
The SM prediction is $BR^{\rm SM}
\approx~4\times 10^{-7}$, and
thus these effects are potentially 
significant departures from the SM.

\newpage
\vskip 0.1in
\noindent
{\em\bf 2.~ $B\to X_{s}\; 
\ell^+\ell^-$}
\vskip 0.1in
The dilepton mass distribution 
has the form
\by
\frac{d\Gamma}{ds}&=&
KF(\q')^2(1-s)^2\sqrt{1-\frac{4x}{s}}
\left\{\left(|C_8|^2+|C'_8|^2-|C_9|^2-|C'_9|^2\right)
6x\right. \nn\\
&
&+\left(|C_8|^2+|C'_8|^2+|C_9|^2+
|C'_9|^2\right)\left[(s-4x)+
(\left(1+\frac{2x}{s}\right)(1+s)\right] \nn \\
& &\left.12C_7{\rm
Re}[C_8-C'_8]\left(1+\frac{2x}{s}\right)+
\frac{4|C_7|^2}{s}\left(1+\frac{2x}{s}\right)
(2+s)\right\} \label{dil_dist}
\ey
where $s=q^2/m_{b}^{2}$ and 
$x=m_{\t}^{2}/m_{b}^{2}$
and where we defined:
\bz
F(\q ')=\frac{2\pi^2\; v^2}{M_{Z'}^{2}}\; 
\left(\frac{m_Z}{m_W}\right)^2 X^2(\q
'); \qquad
K=\frac{G_{F}^{2}\a^2}{768\pi^5}
m_{b}^{5}|V_{tb}V_{ts}^{*}|^2 \label{fac_k}
\ez
and:
\by
C_{8}^{TC}(m_W)&=&-\frac{1}{2}
\frac{D_{L}^{bs}D_{L}^{bb*}}{V_{tb}V_{ts}^{*}} \;
\qquad \qquad
C_{9}^{TC}(m_W)=-\frac{1}{6}
\frac{D_{L}^{bs}D_{L}^{bb*}}{V_{tb}V_{ts}^{*}} \;
\\
C_{8}^{'TC}(m_W)&=&+1
\frac{D_{R}^{bs}D_{R}^{bb*}}{V_{tb}V_{ts}^{*}}\; 
\qquad \qquad
C_{9}^{'TC}(m_W)=+
\frac{1}{3}\frac{D_{R}^{bs}D_{R}^{bb*}}{V_{tb}V_{ts}^{*}}\;
\ey
The Topcolor effect can be large 
for $\tau$ leptons due to the presence of
$X(\q')=\cot\q '$.
To illustrate the possible size of 
the effect we choose again the $\sqrt{\rm
CKM}$ ansatz. In this case we have 
also to choose the sign of $D_{L,R}^{bs}$,
which is taken to be positive. 
For $k_1\sim{\cal O}(1)$ the results are given
in Table~I. The branching ratios 
are similar to those for  $B_s\to
\tau^+\tau^-$, given that the 
partial helicity suppression is balanced by the
space phase suppression in the 
three body decay. There are no presently
published limits on any of 
the $\tau$ channels.
\begin{table}
\centering
\begin{tabular}{|c|c|c|}
\hline
$M_{Z'}[{\rm GeV}]$&$BR(b\to 
s\tau^+\tau^-)$&$BR(b\to s\mu^+\mu^-)$ \\ \hline
$500$&$2.6\times 10^{-3}$&$8.6\times 10^{-6}$\\
$1000$ &$1.6\times 10^{-4}$&$6.0\times 10^{-6}$\\
SM &$4\times 10^{-7}$&$5.7\times 10^{-6}$\\
\hline
\end{tabular}
\vskip 0.75truecm
\caption{Estimates of inclusive 
branching ratios for $b\to s\ell^+\ell^-$ in
the SM and Topcolor. }
\end{table}

\vskip 0.1in
\newpage
\noindent
{\em\bf 3.~ $B\to X_s \nu\bar{\nu}$}
\vskip 0.1in
The decay $b\to s\ell^+\ell^-$ 
could have an important contribution from the
$\tau$ neutrino in Topcolor 
models given that, in principle, they couple
strongly to the $U(1)_1$. The 
Topcolor amplitude can be written as
\bz
{\cal A}^{TC}(b\to s\nu\bar{\nu})=
\frac{g_{1}^{2}}{4}\frac{\cot\q 'X(\q
')}{M_{Z'}^{2}}\left\{ g_v\;\bar{s}
\g_\m b+g_a\;\bar{s}\g_\m\g_5
b\right\}\;\bar{\nu}\g^\m\nu_L 
\label{bsnunu}
\ez
where
\by
g_v&=&\frac{1}{6}\left(D_{L}^{bb}
D_{L}^{bs*}-2D_{R}^{bb}D_{R}^{bs*}\right)
\nn\\
g_a&=&-\frac{1}{6}\left(D_{L}^{bb}
D_{L}^{bs*}+2D_{R}^{bb}D_{R}^{bs*}\right) \nn
\ey
We take the neutral mixing to 
be:
\bz
\d_{bs}= D_{L}^{bb}D_{L}^{bs*}=
D_{R}^{bb}D_{R}^{bs*}\sim \frac{1}{2}\lambda^2
\label{nn_dbs}
\ez
An estimate
of the ratio of branching ratios 
is then:\cite{Burdman}
\bz
\frac{BR^{TC}(b\to s\bar{\n_\t}\n_t)}{BR^{SM}
(b\to s\bar{\n}\n)}\sim
\left\{
\begin{array}{c}
176 k_{1}^{2}  \;\;\; {\rm for }\;\; 
m_{Z'}=\; 500 {\rm GeV} \\
\;\; 11 k_{1}^{2}  \;\;\; {\rm for }\;\; 
m_{Z'}=1000 {\rm GeV}
\end{array}
\right.
\ez
Estimates of this mode in the SM  give 
$BR^{SM}\simeq 7\times 10^{-5}$. 
\vskip 0.1in

\noindent
{\em\bf 4.~ $K^+\to \pi^+ \nu\bar{\nu}$}
\vskip 0.1in
As in $B\to X_s \nu\bar{\nu}$, we are 
concerned with the contact term involving
$\tau$ neutrinos given that they 
constitute the most important Topcolor
contribution. The Topcolor amplitude 
is given by
\bz
{\cal A}^{TC}(K^+\to \pi^+ 
\nu_\t\bar{\nu}_\t) = -\frac{g_{1}^{2}\cot^2\q
'}{24M_{Z'}^{2}} \; \d_{ds} 
f_+(q^2)(p+k)_\mu\;\;\bar{\n}_{\t L}
\g^\m\nu_{\t L} \label{atc_nunu}
\ez
where now $\d_{ds}^{*}=D_{L}^{bs}
D_{L}^{bd*}-2D_{R}^{bs}D_{R}^{bd*}$. The
form-factor $f_+(q^2)$ is 
experimentally well known. 
For one neutrino species
this is given by
\bz
{\cal
A}^{SM}=\frac{G_F\a}{\sqrt{2}
\pi\sin^2\theta_W}\;\; f_+(q^2)(p+k)_\mu
\;\bar{\n}_L\g^\m\n_L\;\;\sum_{j}
V^{*}_{js}V_{jd} D_j(x_j)
\label{sm_amp}
\ez
where $x_j=m_{j}^{2}/m^{2}_{W}$ and $D(x_j)$ 
is an Inami-Lim function. Since
only the vector quark current 
contributes to the exclusive transition the Dirac
structure in the Topcolor and 
SM amplitudes is the same.
The ratio of the Topcolor 
amplitude to the SM is then
\bz
\frac{{\cal A}^{TC}}{{\cal A}^{SM}}=
-\left(\frac{g_1\cot\q
'}{M_{Z'}}\right)^2\;\frac{\sqrt{2}
\pi\sin^2\theta_W}{24\a\; G_F}\;
\frac{\d_{ds}}{\sum_{j}V^{*}_{js}
V_{jd} D_j(x_j)} \label{ratio}
\ez
For $m_t=175$~GeV, $S\approx 
2\times 10^{-3}$. The ratio can be expressed as
\bz
\frac{{\cal A}^{TC}}{{\cal A}^{SM}}=
-3\times 10^9\; \d_{ds}\;
\frac{\kappa_1}{M_{Z'}^{2}} \label{ratio_2}
\ez
The $\sqrt{CKM}$ ansatz yields
\bz
\d_{ds} =-\frac{1}{4}\lambda^5\;(\frac{3}{4}\lambda^5) \label{mix}
\ez
when choosing positive (negative) 
relative signs between the two terms entering
in $\d_{ds}$.
For $M_{Z'}=500$~GeV and  $\kappa_1=1$
the ratio of amplitudes is about $3/2\; (9/2)$. 
Therefore we must worry about
the possible interference. After squaring 
and dividing by $N_\n$ the total
branching ratio is between $1/3 \; (16/3)$ or  $4/3 \; (25/3)$ of the SM one,
depending on the sign of the interference.

\subsection{~Nonleptonic Processes} 

At low energies the Topcolor interactions 
will induce four quark operators
leading to nonleptonic processes.
In Topcolor~I there are potentially large
corrections to $B_d$ and $B_s$ mixing.
There are also
transitions not present in the SM, 
most notably $b\to ss\bar{d}$,
although with very small branching ratios.
Perhaps the most interesting process is
$D^0-\bar{D}^0$ mixing given it is 
extremely suppressed in the SM.
At the charm quark mass scale the dominant 
effect in flavor changing neutral
currents is due to the flavor changing 
couplings of top-pions. In the case of
Topcolor~I the operator inducing  
$D^0-\bar{D}^0$ mixing can be written as
\bz
\fr{m_t}{\sqrt{2}f_{\tilde{\pi}}}\,
\d_{cu}\,\bar{u}\g_5c\,\bar{u}\g_5c
\label{cu_coup}
\ez
where $\d_{cu}$ is the factor arising 
from the rotation to the mass
eigenstates.
The contribution of (\ref{cu_coup}) 
to the mass difference takes 
the form
\bz
\D m_{D}^{TCI}=\fr{5}{12}\, f_{D}^{2} m_D\,
\fr{m_{t}^{2}}{f_{\tilde{\p}}^{2}
m_{\tilde{\p}^0}^{2}}\, \d_{cu} 
\label{dd_mix}
\ez
where $f_D$ is the $D$ meson 
decay constant.
In the $\sqrt{\rm CKM}$ ansatz 
and for a top-pion mass of
$m_{\tilde{\p}^0}=200$~GeV we 
obtain
\bz
\D m_{D}^{TCI}\approx 2\times 
10^{-14} {\rm~GeV} \label{tcdm}
\ez
which is approximately a factor of 
five  below the current experimental limit.
On the other hand, the SM predicts 
$\D m_{D}^{SM}<10^{-15} {\rm~GeV}$. This
puts potentially large Topcolor 
effects in the discovery window of future high
statistics charm experiments!

The effect could be  even stronger 
in Topcolor~II.\cite{Burdman} 
In the $\sqrt{\rm CKM}$ ansatz this 
gives
\bz
\D m_{D}^{TCII}\approx 4\times 10^{-12}~{\rm GeV}
\ez
which violates the current experimental 
upper limit by about an order of
magnitude. This is the single most 
constraining piece of phenomenology on this
model. It can be avoided by taking 
a different ansatz for the matrices $U_L$
and $U_R$. It would not be natural 
to have no off-diagonal elements in $U_L$
given that it is one the factors in 
the CKM matrix. However $U_R$ could be
almost diagonal in which case the 
effect would be largely suppressed.
The same is true in Topcolor~I. 
Triangular textures in the up sector
are an example of how this can 
be achieved.
 Very large effects in  
$B^0-\bar{B}^0$
mixing are also avoided 
if triangular textures are 
present in the down sector. 

\subsection{~High Energy Processes} 

There are other areas in which 
Topcolor manifests itself at higher
energies, but below the Topcolor 
threshold. For example, there will
be effects in $Z\rightarrow b\bar{b}$. 
Initially Zhang and Hill
considered only topgluons and found 
positive enhancements of $\Delta R_b$.
However, a more thorough analysis of 
Burdman and Kominis finds that the
top-pion contributions  are dominant 
and yield a negative $\Delta R_b$,
and may impose considerable constraints 
on the theory \cite{Burdman2}.
Their result is interesting, but we 
feel not sufficiently
robust for reasonable uncertainties 
in the strong dynamics. For example,
for $f_\pi\sim 60$ GeV almost all 
reasonable top-pion masses are
excluded, however for $f_\pi\sim 120$ 
GeV all top-pion masses are
acceptable.  We do not necessarily 
propose a large $f_\pi$, but this
illustrates the extreme sensitivity 
of the result of \cite{Burdman2}
to corrections of order unity.

There are significant constraints 
from the $\rho$-parameter which
have been studied by Chivukula and 
coauthors \cite{Chiv}. These
evidently imply that the topgluon
 mass scale must significantly exceed
$\sim 1$ TeV.  A more
detailed analysis of this in the
 effective Lagrangian of top-pions
(and other low mass boundstates) 
should be undertaken, in analogy
to \cite{Burdman2}.  In my opinion 
this is potentially critical
for these models.

Effects of topgluons
in top production have been discussed
 by Hill and Parke, \cite{HP} 
and preliminary analysis undertaken 
at CDF by Harris.\cite{Harris}
The clear indication of the top-gluon 
would be a peak in the $t\bar{t}$
invariant mass at the pole, much like 
a Drell-Yan peak in $\mu^+\mu-$.
It is possible the Tevatron lacks 
sufficient energy to make the
full resonance, so an excess of 
high mass top pairs would be the
next indication and the LHC would 
be required to produce the topgluon. 
The excess of Hi-p$_T$ jets can be 
explained by topgluons as well, 
though the more conservative explanation 
involves a modified gluon distribution in
the proton.  An interesting variation 
on the topcolor idea,
flavor democratic colorons, has been 
proposed to explian the Hi-p$_T$
excess.\cite{Chiv2}

These will be interesting 
issues in the Tevatron Run II.

\section*{}


\begin{thebibliography}{99} 

\bibitem{RGFP} B. Pendleton, 
G.G. Ross, {\em Phys. Lett.}
{\bf 98B} 291, (1981); C.T. Hill,
{\em Phys. Rev.} {\bf D24}, 691 (1981); 
C.~T. Hill, C.~N. Leung, 
S. Rao, {\em Nucl. Phys.} {\bf B262},  517 (1985).


\bibitem{Carena} see, for example, 
 W. A. Bardeen, M. Carena, S. Pokorski, 
C. E. M. Wagner,
{\em Phys. Lett. } {\bf B320}, 110, (1994). 

\bibitem{BHL}
Y. Nambu, 
``BCS Mechansim, Quasi-Supersymmetry, 
and Fermion
Mass Matrix,'' Talk presented at the Kasimirz
Conference, EFI 88-39 (July 1988);\\
``Quasi-Supersymmetry, Bootstrap 
Symmetry Breaking, and Fermion
Masses,'' EFI 88-62 (August 1988); 
a version of this work appears in
``{\em 1988 International Workshop 
on New Trends in Strong Coupling
Gauge Theories},'' Nagoya, Japan, 
ed. Bando, Muta and Yamawaki;
``Bootstrap Symmetry Breaking in 
Electroweak
Unification,'' EFI Preprint, 89--08 (1989).
V.~A. Miransky, M. Tanabashi, 
K. Yamawaki,
{\em Mod. Phys. Lett.} {\bf A4}, 1043 (1989);
{\em Phys. Lett.} {\bf 221B} 177 
(1989); 
W.~J. Marciano,  {\em Phys. Rev. Lett. } 
{\bf 62}, 2793 (1989).
W.~A. Bardeen, C.~T. Hill, M.~Lindner
{\em Phys. Rev.} {\bf D41}, 1647 (1990).

\bibitem{Miransky} R.R. Mendel, 
V.A. Miranskii, {\em Phys. Lett.} 
{\bf B268} 384, (1991); 
V.A. Miranskii, {\em  Int. J. Mod. Phys.} {\bf A6}
1641, (1991). 

\bibitem{TC2}
C.~T. Hill, {\em Phys. Lett.}
{\bf B266} 419 (1991); 
 
\bibitem{TC} S. Weinberg, 
{\em Phys. Rev.}
{\bf D13}, 974 (1976);
L. Susskind, {\em Phys. Rev.}
{\bf D20}, 2619 (1979);
 S. Dimopoulos, L. Susskind,
{\em  Nucl. Phys.} {\bf B155}
237 (1979); E. Eichten, K. Lane,
{\em  Phys. Lett.} {\bf 90B}
125 (1980).

\bibitem{TC22} C.~T. Hill, 
{\em Phys. Lett.}
{\bf B345} 483, (1995).


\bibitem{TC23} S. P. Martin,
{\em Phys. Rev.} {\bf D46}, 2197 (1992);
{\em Phys. Rev.} {\bf D45}, 4283 (1992),
{\em Nucl. Phys.} {\bf B398}, 359 (1993);
M. Lindner and D. Ross, {\em Nucl. Phys.}
{\bf  B 370}, 30 (1992);
 R. B\"{o}nisch, {\em Phys. Lett.}
{\bf B268} 394 (1991);
C.~T. Hill, D. Kennedy,
T. Onogi, H.~L. Yu, {\em Phys. Rev.}
{\bf D47} 2940  (1993).

\bibitem{Lane} K. Lane, E. Eichten, 
{\em Phys. Lett.} {\bf
B352}, 382, (1995); 
 K. Lane, {\em Phys. Rev.} {\bf D54}
2204, (1996).

\bibitem{Burdman}  G. Buchalla, 
G. Burdman, C. T. Hill, D. Kominis, 
{\em Phys. Rev. } {\bf D53} 5185, (1996). 

\bibitem{BB} ``Top Decay in Topcolor 
Assisted Technicolor,''
 B. Balaji, BUHEP-96-29, (1996); 
hep-ph/9610446. 

\bibitem{Burdman2}  G. Burdman,
 D. Kominis, ``Model Independent
 Constraints on Topcolor from
$R_b$,'' MADPH-97-984, TUM-HEP-267/97 (1997), 
hep-ph/9702265;
 C. T. Hill, Xinmin Zhang, 
{\em Phys.Rev. } {\bf D51}
3563, (1995).


\bibitem{Chiv} R. S. Chivukula,
 B. Dobrescu, J. Terning,
 {\em Phys.  Lett.}
{\bf B353}, 289 (1995);
R. S. Chivukula, J. Terning, {\em Phys. Lett.}
{\bf B385} 209 (1996).


\bibitem{HP} C. T. Hill,
S. J. Parke, {\em Phys. Rev.} {\bf D49},
4454   (1994); E. Eichten, K. Lane  {\em Phys. Lett. } {\bf B388}
803 (1996).


\bibitem{Harris} R. M. Harris, 
FERMILAB-CONF-96-276-E, (1996);
FERMILAB-CONF-96-277-E, (1996).

\bibitem{Chiv2} R.S. Chivukula, 
A.G. Cohen, E.H. Simmons 
{\em Phys. Lett. } {\bf B380} 
92 (1996). 
E. H. Simmons {\em Phys. Rev. } 
{\bf D55} 1678 (1997). 

\end{thebibliography}
\end{document}